\documentclass[aps,onecolumn,superscriptaddress,super]{revtex4-2}
\usepackage{graphicx,color,setspace}
\usepackage[utf8]{inputenc}
\usepackage{mathtools}
\usepackage{tabularx}
\usepackage{enumerate}%
\usepackage{float}
\usepackage{ragged2e} %
\usepackage{subcaption}
\usepackage{setspace}
\makeatletter
\renewcommand{\cite}[1]{\textsuperscript{\citenum{#1}}}
\makeatother
\DeclareCaptionJustification{justified}{\justifying}

\usepackage[labelfont=bf]{caption}
\begin{document}

\title{A single-chain nanoparticle-based mean-field theory for associative polymers}
\author{Marco Cappa}
\email[The author to whom correspondence should be addressed, ]{marco.cappa@uniroma1.it}
\affiliation{Dipartimento di Fisica, Sapienza Universit\`{a} di Roma, P.le Aldo Moro 5, 00185 Rome, Italy}

\author{Stefano Chiani}
\affiliation{Department of Physics, University of Ioannina, P. O. Box 1186, 451 10 Ioannina, Greece}

\author{Francesco Sciortino}
\affiliation{Dipartimento di Fisica, Sapienza Universit\`{a} di Roma, P.le Aldo Moro 5, 00185 Rome, Italy}

\author{Lorenzo Rovigatti}
\affiliation{Dipartimento di Fisica, Sapienza Universit\`{a} di Roma, P.le Aldo Moro 5, 00185 Rome, Italy}
\date{\today}

\begin{figure}[H]
    \centering
    \includegraphics{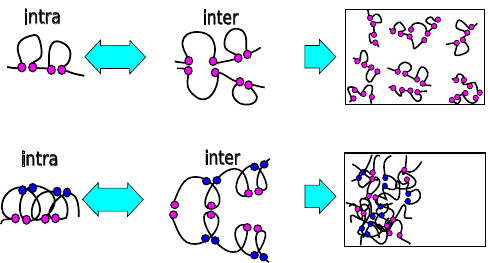}
    \label{fig:toc}
\end{figure}

\begin{abstract}
\noindent\textbf{Abstract:}
Associative polymers are a class of polymers containing attractive stickers that can reversibly bind to each other. Their fully-bonded state gives rise, in dilute conditions, to a fluid phase of so-called single-chain nanoparticles (SCNPs). These constructs have been used in a wide range of applications, from the design of new materials (\textit{e.g.} biomolecular condensates) to drug delivery vectors.  The thermodynamic properties of SCNPs sensitively depend on the number of different sticker types, since numerical simulations show that a continuous transition to a network of chains upon increase of polymer concentration in the single sticker-type case can be replaced by an abrupt network formation (via a first-order phase transition) in the multiple sticker-type case. 
We present here a theory that, using the SCNP fluid as the reference system, quantifies the free energy change associated with transferring an intra-molecular bond to an inter-molecular bond, elucidating
the impact on the  phase separation process of the sticker topology.  Despite its simplicity, the theory highlights which microscopic assumptions (loop statistics, chain-level excluded volume) are most relevant for accurately capturing the thermodynamics of these systems. Our results match available numerical predictions obtained via coarse grained simulations of these systems, highlighting in particular the sensitivity of the phase behaviour on the sequence of the stickers along the chain. 
\end{abstract}

\maketitle

\section{Introduction}
Polymers are a class of macromolecules with unique properties: their large conformational freedom, combined with their periodic chemical structure, make them an interesting subject of study in many contexts, from statistical mechanics to materials science and biophysics. Relevant applications include, among others, their use in the design of new synthetic materials~\cite{fattah2024, wang2020, das2022, gonzalez2021, xiao2020}, as drug delivery vectors~\cite{bailey2021, matange2025}, or as models for many biological phenomena~\cite{ranganathan2020dynamic,borandeh2021, salari2022}.

While extensive research has been dedicated to the description of regular (\textit{i.e.} so-called \textit{athermal}) polymer solutions, especially with respect to their equilibrium properties and thermodynamic behaviour~\cite{flory1941, huggins1942}, comparatively less effort has been devoted to the study of \textit{associative polymers}, \textit{i.e.}, chains in which special reactive monomers (named \textit{stickers}) are distributed along the chain, separated by a number of chemically inert (spacer) monomers~\cite{semenov1998, friedrickson_2009, choi2020, ranganathan2020dynamic, danielsen_2023, delacruz_2024, riggleman_2025,chen_2026}. Usually, spacer monomers are modelled as particles interacting solely through steric repulsion, whereas the much less abundant stickers are relatively small, strongly aggregating functional groups of potentially different types. When stickers are close enough, they are able to form bonds with other stickers through a wide range of interactions, ranging from metal complexation~\cite{mahmad2020} and ionic or hydrophobic interactions~\cite{abdala2003} to hydrogen~\cite{feldman2009, cordier2008} and even covalent~\cite{denissen2016, montarnal2011} bonds. In many cases, each sticker can be involved in no more than a single bond and, in the case of multiple sticker types, only with a sticker of a compatible type. Throughout the years, these foregoing ingredients have been used to develop models that have been successfully applied to several contexts, from biomolecular condensates~\cite{choi_physical_2020} and RNA-protein mixtures~\cite{chen_solgel_2025}, to vitrimers~\cite{smallenburg2013patchy,rovigatti2018self,xia_structure_2023} and single-chain nanoparticles (SCNPs), which are self-folding macromolecules that are of potential interest in a wide range of applications~\cite{pomposo2015, rothfuss2018, moreno2018, verdesesto2020}.

When a bond between two stickers forms, the interacting pair of monomers is constrained to move together, thus reducing the conformational entropy of the system to an extent that depends on whether the bond is \textit{intermolecular} (\textit{i.e.}, between stickers belonging to different chains) or \textit{intramolecular} (\textit{i.e.}, between stickers within the same chain)~\cite{semenov1998}. This entropic term, intrinsic to the polymeric nature of the particles, plays a major role in determining the thermodynamics of associative polymers. The dependence on the number and type of attractive sites of this conformational entropy contribution is complex~\cite{moreno2018}, yet crucial to correctly predict the thermodynamic properties of the system.

For associative polymers with a single type of stickers in a good solvent, theory (in the limit of strongly aggregating stickers) and experiments have shown that, upon increasing the concentration, a solution of chains in a good solvent forms a network continuously, without signs of a thermodynamic phase transition~\cite{semenov1998,whitaker_thermoresponsive_2013,chen_solgel_2025,ruiz2025,rovigatti2022designing}. By contrast, it was shown by computer simulations that, if the stickers of the SCNPs along the chain are of two or more different and alternating types, in the fully-bonded limit (\textit{i.e.} large monomer-monomer attraction strength compared to the thermal energy) a significant entropic contribution induces a first-order phase transition: the network forms abruptly at a certain concentration~\cite{rovigatti2023,rovigatti2022designing}. Indeed, changing the number of types of reactive monomers affects the ratio of intra- to intermolecular bonds and thus the overall inter-polymer large-scale connectivity. This behaviour has been recently confirmed by realistic simulations of all-DNA-based SCNPs~\cite{tosti_guerra_entropy-driven_2025}.

The appearance of such a phase transition can be understood by noting that, starting from a chain decorated by stickers of the same type and everything else being equal, converting some of them to a different type de facto increases the average distance between identical stickers, \textit{i.e.} the contour length of the loop between bonded stickers.  The resulting fully-bonded SCNP will have a smaller gyration radius. As a further consequence, we envision that there will be a greater entropic gain from swapping intramolecular for intermolecular bonds.  We show here that if such a gain is large enough, the system will phase-separate even in a good solvent. To quantify these expectations, we introduce a mean-field expression for the free energy cost of swapping intra-molecular to inter-molecular bonds, estimating how such a cost depends on the number of different sticker types. Adding such free-energy cost of swapping  to the reference system free energy (which is estimated up to the second term of the SCNP virial expansion)  provides predictions for the phase behaviour that recapitulate the numerical results, providing support to the idea of leveraging the sticker sequence to manipulate the thermodynamics of associative polymers. 

\section{A mean-field theory for sticker association}

\begin{figure}
    \centering
    \includegraphics[width=0.5\linewidth]{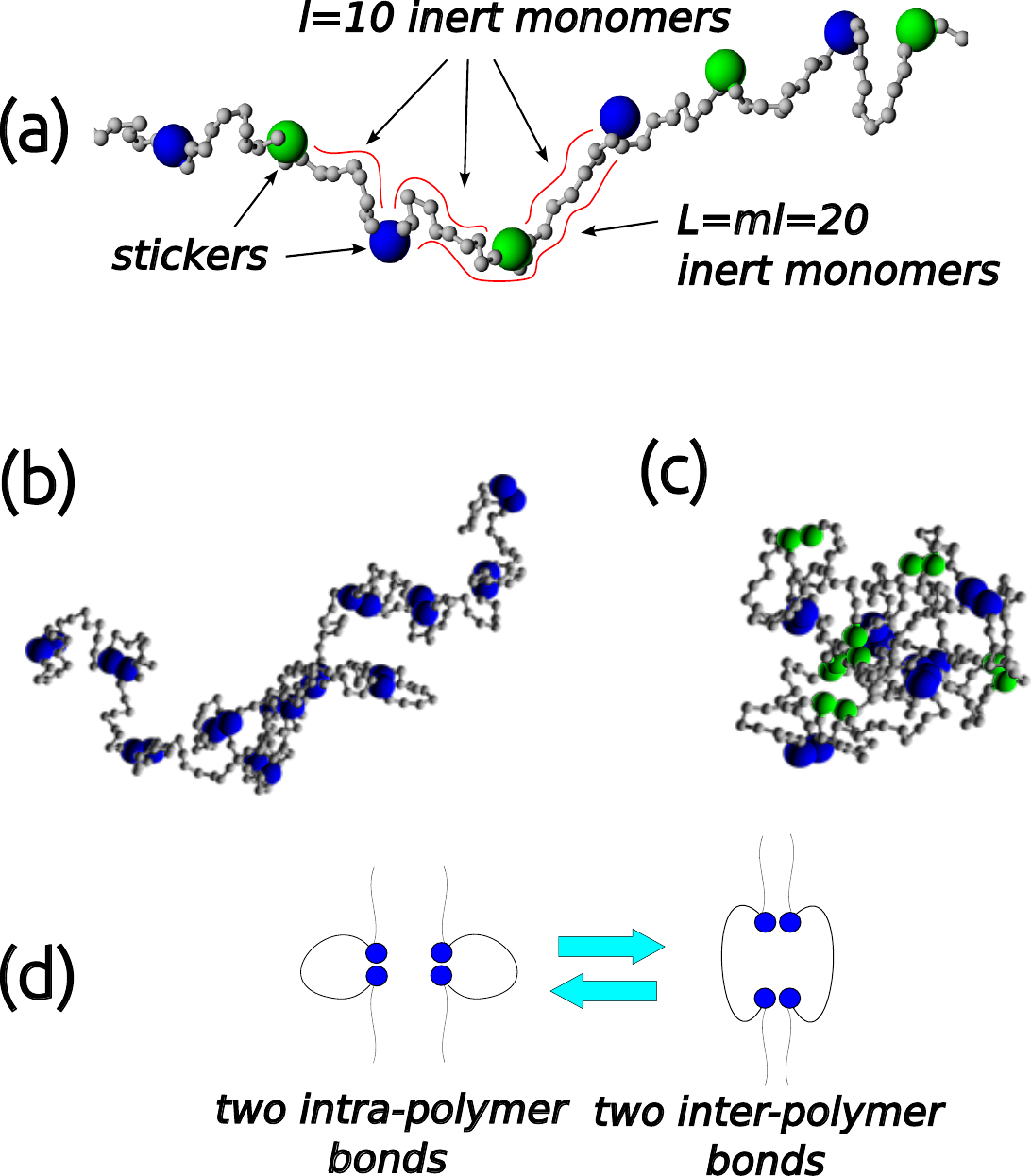}
    \caption{(a) Segment of an associative polymer chain with $l=10$, $m=2$ and $L=ml=20$. (b-c) Fully-bonded configurations of (b) $m = 1$ and (c) $m = 2$ associative polymers with $\mathcal{M} = 24$ stickers and $l = 10$ inert monomers between stickers. The $m=2$ polymer presents a more compact structure because of the alternating looping scheme. Note that in the (a-b-c) panels, for clarity, inert monomers and stickers are shown with smaller and larger diameters, respectively, than their actual physical size. (d) Cartoon of the ``chemical reaction'' between two $A_{i1}$ units (forming two intra-polymer bonds) and one $A_{i2}$ unit (forming two inter-polymer bonds).}
    \label{fig:schema}
\end{figure}

We consider a solution of associative polymers consisting of chains made by $N$ monomers of size $a$, which in the following will be used as unit of length. Each chain bears $\mathcal{M} \ll N$ equally-spaced associative groups of $m$ different types (the \textit{stickers}) which alternate along the backbone. Two consecutive stickers are separated by $l=\frac{N}{\mathcal{M}}$ inert monomers, so that two consecutive stickers of the same type are separated by $L\approx m\frac{N}{\mathcal{M}}=ml$ such monomers, as shown in Fig.~\ref{fig:schema}(a). Inert monomers interact solely through steric repulsion, while each sticker is able to form a single reversible bond with another sticker of the same type. We are interested in developing a theoretical characterisation of the thermodynamic behaviour of the system in the limit in which essentially all the stickers are involved in a bond. Such a fully-bonded state can be achieved by either considering the low-temperature limit, or the large bond-strength limit. Since all possible bonds are formed, the total number of bonds is fixed, with individual bonds switching from inter-chain to intra-chain and vice versa. Under these conditions, to a first approximation the system internal energy (which is dominated by the bond contribution) remains constant regardless of the way bonds are distributed among the polymers, so that the equilibrium properties of the system are controlled by the maximisation of the total entropy.

In the fully-bonded limit, under highly diluted conditions, all bonds are intra-chain. Each polymer constitutes aFn independent entity, which we call a SCNP.
For fully-flexible polymers, the entropic cost of binding (which forces the inert monomers between the bonded sites to form a closed loop)  scales with the size of the loop, and therefore intramolecular bonds tend to occur between consecutive stickers of the same kind~\cite{semenov1998}. If we let the number of stickers of each type be even, the chains fold on themselves in closed, cactus-like structures.
On increasing the number of different stickers, the fully-bonded structure becomes increasingly compact, as shown in panels (b) and (c) of Fig.~\ref{fig:schema}. This solution of weakly interacting fully-bonded SCNPs constitutes the \textit{reference} system with respect to which inter-chain bonding contributions are considered. 

We write the Helmholtz free energy density $\beta f$ (here $\beta=\frac{1}{k_B T}$, with $T$ indicating the temperature of the system and $k_B$ the Boltzmann constant) of the system using a mean-field perturbative approach as:
\begin{equation}
\beta f(c)=\beta f_{\rm {ref}}(c)+\beta f_{\rm bond}(c)
\label{eq:freee}
\end{equation}
where $c$ is the number density of monomers. The first term $\beta f_{\rm ref}$ is the free energy of a system of independent SCNPs and constitutes the sum of an ideal-gas-of-chains term $\beta f_{\rm ig }(c)=\frac{c}{N}\log{\frac{c}{N e}}$ (note that $e$ is Napier's number and $\frac{c}{N}$ is the number density of \textit{chains}) and a purely repulsive, steric contribution $\beta f_{\rm ex}$.  The latter is modelled as a first correction to the ideal gas, i.e.
\begin{equation}
\beta f_{\rm ex}(c)=B_2 \frac{c^2}{N^2}
\label{eq:b2}
\end{equation}
where $B_2$ is the (purely repulsive) second virial coefficient of the fully-bonded SCNP.

\begin{figure}
    \centering
    \includegraphics[width=0.6\linewidth]{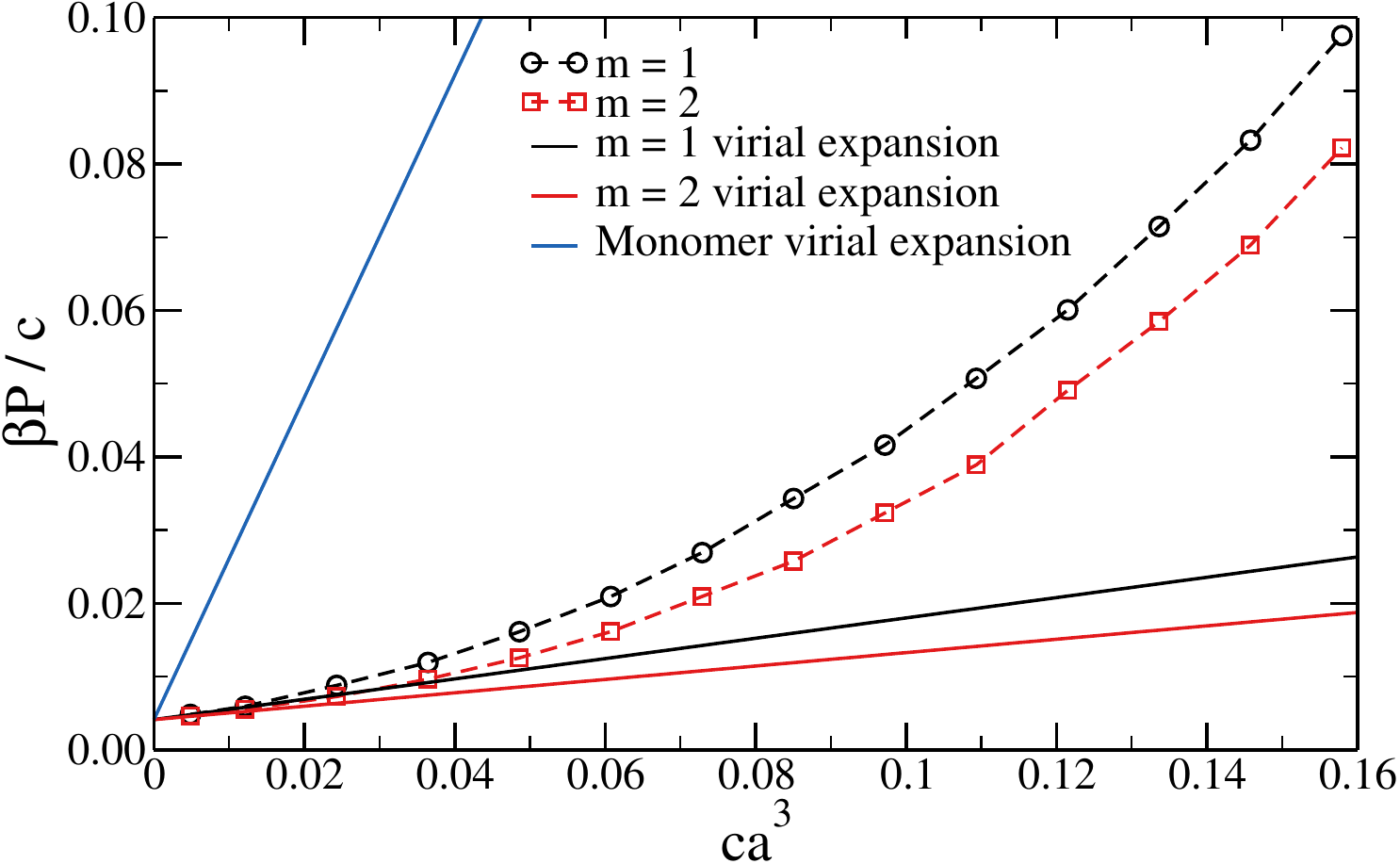}
    \caption{The compressibility factor of fully-bonded systems made of $m = 1$ and $m = 2$ chains, when only intra-chain bonding is allowed (dashed lines). The solid lines have been drawn by considering a second-order virial expansion, fixing the value of the $B_2$ to that extracted from independent two-body effective interaction simulations (black and red lines, see text for details), or to the second virial coefficient of single monomers, $B_2^{\rm mon} \approx 2.2 a^3$.}
    \label{fig:B2_pressure}
\end{figure}

To assess the quality of the approximated reference free energy (Eq.~\ref{eq:b2}), we perform simulations of a system of fully intra-bonded SCNPs with one ($m=1$) or two ($m=2$) types of stickers in the limit of large binding strength. Fig.~\ref{fig:B2_pressure} compares the equation of state (EOS) of fully intra-bonded SCNPs with the analytical second-order virial expansion in the chain number density, $c/N$, using the $B_2$ value extracted from independent two-body effective interaction simulations (see the~\textit{Methods}). Up to $ca^3 \approx 0.04$, a value rather close to the critical density of these systems~\cite{rovigatti2023}, the virial expression properly reproduces the EOS.
By contrast, a second-order virial expansion at the monomer level (\textit{i.e.} in the monomer density $c$), which is what has been often used in the past~\cite{semenov1998}, appears to overestimate (in the case of a good solvent) the excess contribution to the pressure of the system. It is worth noting that alternative approaches, based on choosing the optimal value of the monomer excluded volume to maximise the agreement between theoretical predictions and simulation-derived phase diagrams, can also be employed~\cite{chen_2026}.

The second term in Eq.~\ref{eq:freee} expresses the change in the system when intra-polymer bonds are swapped for inter-polymer bonds. We  thus focus  on the inter-polymer bonds and on the association phenomenon resulting from the swap between intra- and inter- bonds.  

Specifically, we envision two bonded stickers of the same type $i$ on the same polymer as a chemical species $A_{i1}$, or equivalently a 
complex of inert monomers separated by two identical stickers. By a bond-swapping process, two $A_{i1}$ units, each on a different polymer, can react to form a $A_{i2}$ intermolecular complex, as illustrated schematically in Fig.~\ref{fig:schema}(d). Therefore, the process can be formally modelled as a chemical reaction:
\begin{equation}
\label{eq:chemical_eq}
A_{i1}+ A_{i1} \rightleftharpoons A_{i2}
\end{equation}

In a more precise thermodynamic formalism, the chemical equilibrium can be written as (indicating with $N_{A_{i1}}$ and $N_{A_{i2}}$
the number of $A_{i1}$ and $A_{i2}$ units and with 
$Q_{A_{i1}}$ and $Q_{A_{i2}}$ their partition functions, respectively~\cite{hill2012introduction,sciortino2016}):
\begin{equation}\label{eq:mass_action}
N_{A_{i2}} = Q_{A_{i2}} \left(\frac{N_{A_{i1}}}{Q_{A_{i1}}}\right)^2
\end{equation}
The expressions for $Q_{A_{i1}}$ and $Q_{A_{i2}}$ are reported in Appendix~\ref{app:mass_action}.  In terms of the \textit{degree of conversion} $q_i=\frac{N_{A_{i2}}}{N_i/4}$ (the fraction of $A_{i2}$ units formed from the total number of stickers of type $i$, $N_i$), the law of mass action corresponding to the chemical reaction of Eq.~\eqref{eq:chemical_eq} is:

\begin{equation}\label{MAL}
\frac{q_i}{(1 - q_i)^2} =  \frac{4c v_L}{2^{3\nu+1}L}
\end{equation}

\noindent
where $v_L$ is the volume accessible to the  chain of length $L$ separating two consecutive stickers, which we assume to scale as $v_L=a^3 L^{3\nu}$, and $\nu$ is a scaling exponent which will be discussed further below. Since the right-hand side of the previous equation is actually independent of $i$, we can consider all the $\{q^i\}$ to be the same and set them to $q$. Resolving with respect to $q$, we obtain:

\begin{equation}\label{q}
q=1-\frac{-1+\sqrt{1+ \frac{16cv_L}{2^{3\nu+1}L}}}{\frac{8 c v_L}{2^{3\nu+1}L}}
\end{equation}

It is necessary to point out that this approach assumes an uncorrelated degree of conversion $q$, \textit{i.e.} it neglects correlations and cooperative behaviour in sticker-sticker interactions, coherently with a mean-field hypothesis. More sophisticated theories incorporating bonding correlations (see \textit{e.g.} Ref.~\cite{choi_2020}) have to be used when investigating systems dominated by such effects, which can be brought about by \textit{e.g.} electrostatic interactions involving charged groups in polyampholytes~\cite{mccarty_2019}.

\begin{figure}
    \centering
    \includegraphics[width=0.7\linewidth]{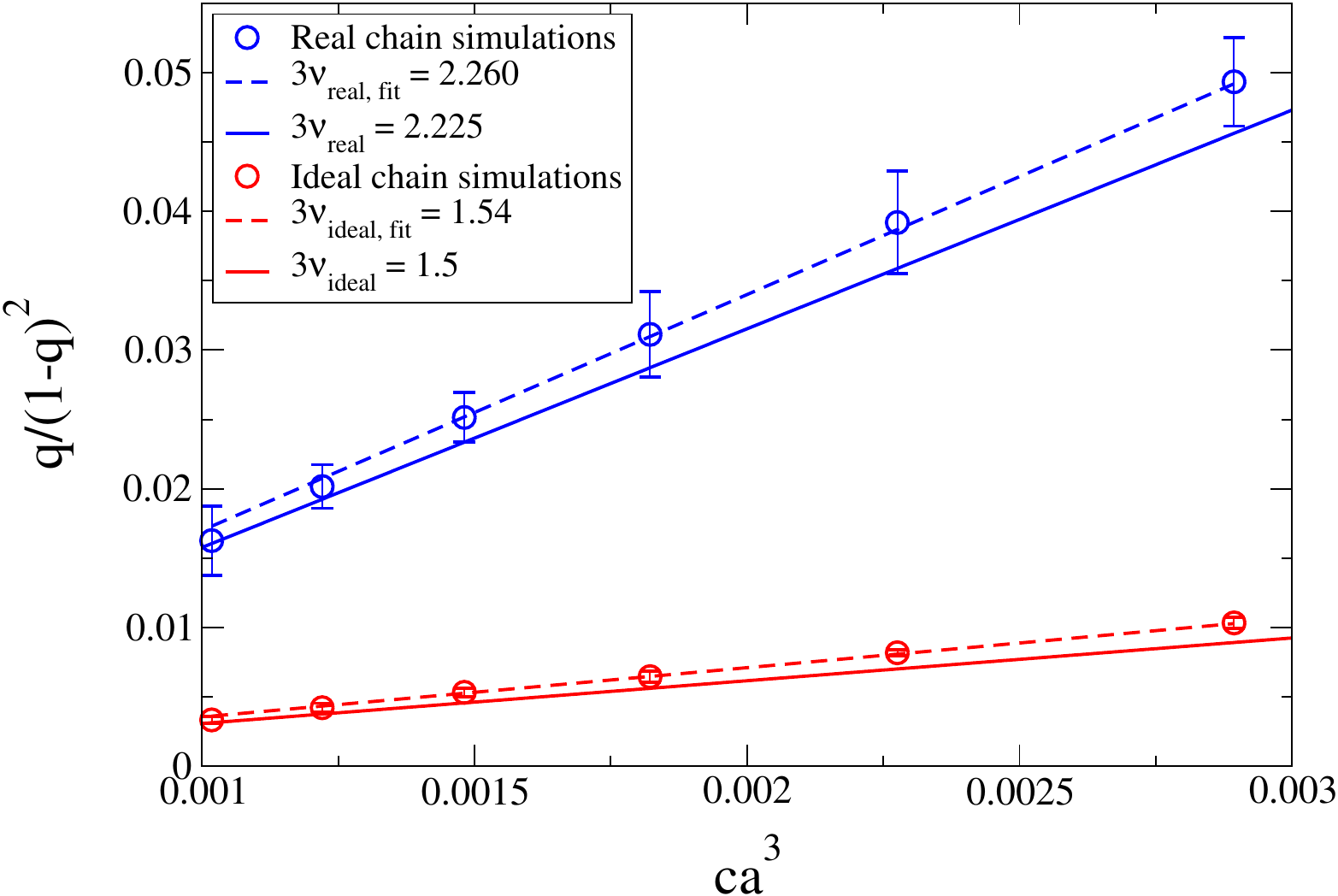}
    \caption{Mass action laws from MD simulations of a polymeric sequence  with $l=18$ inert monomers and two additional stickers, one at each end. To reproduce the ideal chain behaviour, the non-bonded monomer-monomer interaction has been set to zero.  
    The values shown have been computed running 8 simulations for each state point and calculating the average value $\langle q/(1-q)^2 \rangle$ for each state point. The error bars correspond to the standard error. Dashed lines are one-parameter fits to Eq.~\eqref{MAL}, which yield $3\nu_{\textrm{real}}=2.25$ and $3\nu_{\textrm{ideal}}=1.56$, while solid lines are parameter-free curves obtained by plugging the theoretical values $3\nu^*_\textrm{real}=2.225$ and $3\nu^*_{\textrm{ideal}}=1.5$ in Eq.~\eqref{MAL}.   In these calculations we have chosen $\beta \epsilon=6$ ($\epsilon$ indicates the attractive potential depth). Only fully-bonded configurations have been included in the analysis. 
     }
    \label{fig:mass_action}
\end{figure}

To check the accuracy of the estimate of the partition functions of the $A_{i1}$ and $A_{i2}$ units, we carry out MD simulations of two $A_{{i1}}$ units
(modelled as polymers of 20 monomers with sticky sites at both ends) and compute $\frac{q}{(1-q)^2}$ (defined in this case as the fraction of inter-chain-bonded configurations out of all the fully-bonded configurations obtained) as a function of the monomer concentration $c$. We calculate $\frac{q}{(1-q)^2}$ both including and excluding the repulsive interactions between non-adjacent monomers, to test the relation for ideal and self-avoiding polymers.  A thorough explanation on how these simulations were carried out can be found in the~\textit{Methods}.
The results of these computations are presented in Fig.~\ref{fig:mass_action} as symbols.
The line through the data is a power-law fit, consistent with 
Eq.~\eqref{MAL}. The best-fit critical exponent 
$\sim L^{3\nu}$  is close to the expectation values from standard polymer theory for both ideal ($3\nu^*_{\textrm{ideal}}=1.5$) and self-avoiding ($3\nu^*_{\textrm{real}}=2.225$) chains,  confirming the reliability of our estimate of the inter-chain bonding contribution. It is worth noting that here the expected real chain exponent is higher than the standard Flory estimate $\nu^*_{\textrm{flory}}=0.588$ because it takes into account the enhancing effect that excluded volume interactions have on the inter-chain bonding (see \textit{e.g.} Ref.~\cite{semenov1998}), which makes the entropic gain upon intra-chain bond breaking even larger.

We can now finalise the bonding contribution to the free-energy, based on the expressions derived to model associating network forming liquids by Wertheim~\cite{wertheim1984fluids}, as (see Appendix~\ref{app:free_en_clusters} for a simple derivation based on
thermodynamic perturbation theory (TPT)~\cite{chapman1988phase}):

\begin{equation}
\beta f_{\rm bond}=\sum_{i=1}^m \frac{N_i}{2V} \left(\frac{q_i}{2}+\log{(1-q_i)}\right)
\label{sticker_free_en}
\end{equation}

Since $q_i = q$ $\forall i$,  such a bonding contribution to the free energy density simplifies to 
\begin{equation}
\beta f_{\rm bond}=m \frac{c}{2L}\left(\frac{q}{2}+\log{(1-q)}\right)
\end{equation}
This expression, combined with the reference free-energy, provides the total free energy density of the model:
\begin{equation}\label{free_energy}
\beta f(c)=\frac{c}{N}\log{\frac{c}{e N}}+ \beta f_{\rm ex}(c) + m \frac{c}{2L}\left( \frac{q}{2} + \log{(1-q)}\right)
\end{equation}

We now note that  $\beta f(c)$ takes, barring a different expression for $\beta f_{\rm{ex}}(c)$ and $q$ and a factor $1/2$ in $\beta f_{\rm{bond}}(c)$, the same structure as the extension of Eq.~(4.29) in the renormalised theory by Semenov and Rubinstein of Ref.~\cite{semenov1998} to the case of multiple sticker types $m>1$, obtained by leveraging a polymer-physics based approach (for a complete derivation see Appendix ~\ref{app:sem_rub_gen}). Since the prefactor $m/L=1/l$ in the bonding contribution is independent of $m$, the enhancing effect on phase separation due to the number of different types of stickers $m$ is solely linked to the larger average length between alike stickers in $v_L$, \textit{i.e.} to the higher entropy unfrozen upon intramolecular loop breaking.

\section{Phase diagram}

The proposed free energy expression, derived in the limit of fully-bonded polymers,  depends on the two parameters $c$ and $B_2$.  In such a plane, it is possible to calculate the binodal line  (similar to the liquid-gas phase separation) and the critical point. More detailed explanations on the phase diagram calculation procedure are reported in the~\textit{Methods}.

\begin{figure}
    \centering
    \includegraphics[width=0.6\linewidth]{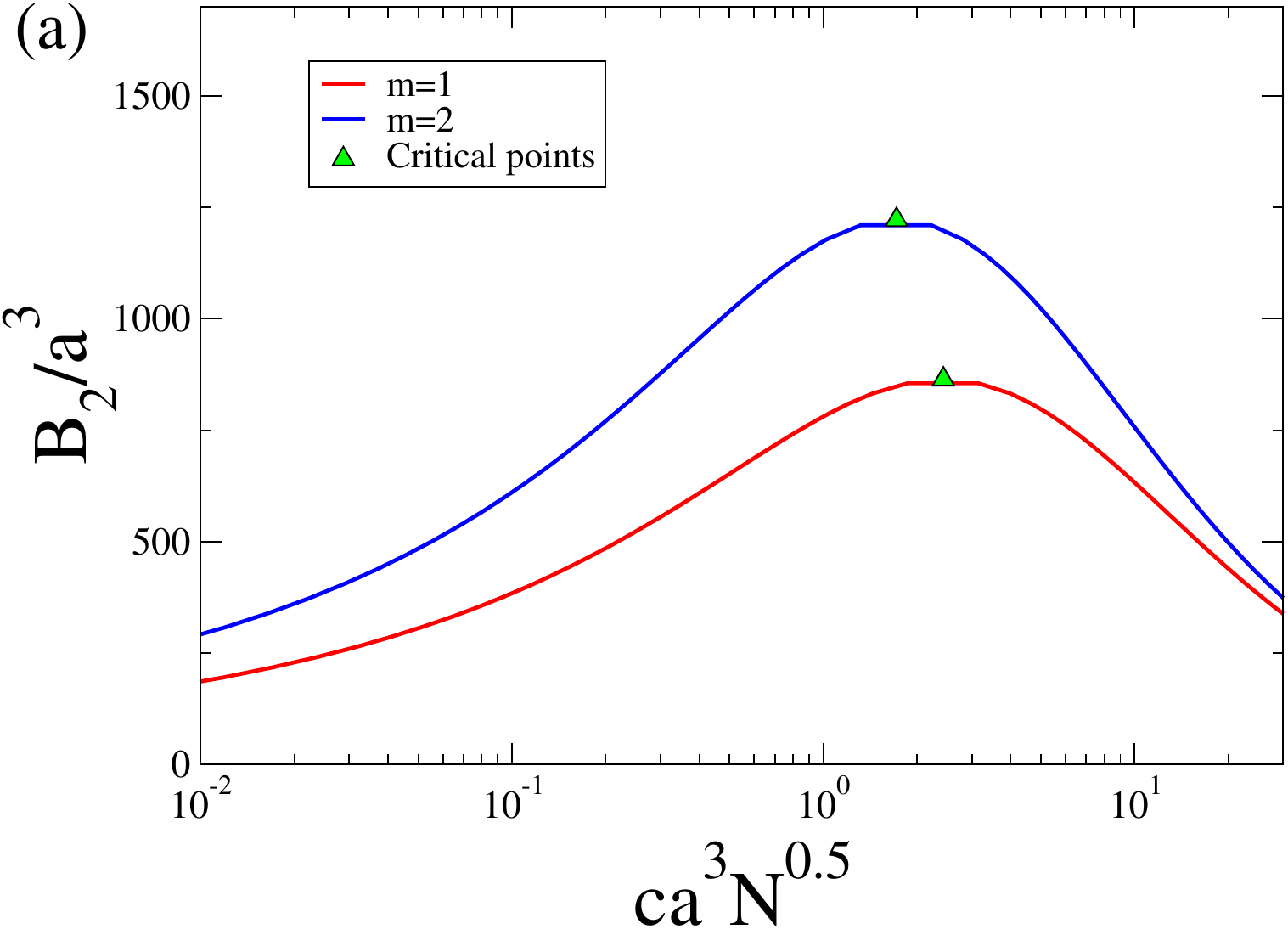}
    \includegraphics[width=0.6\linewidth]{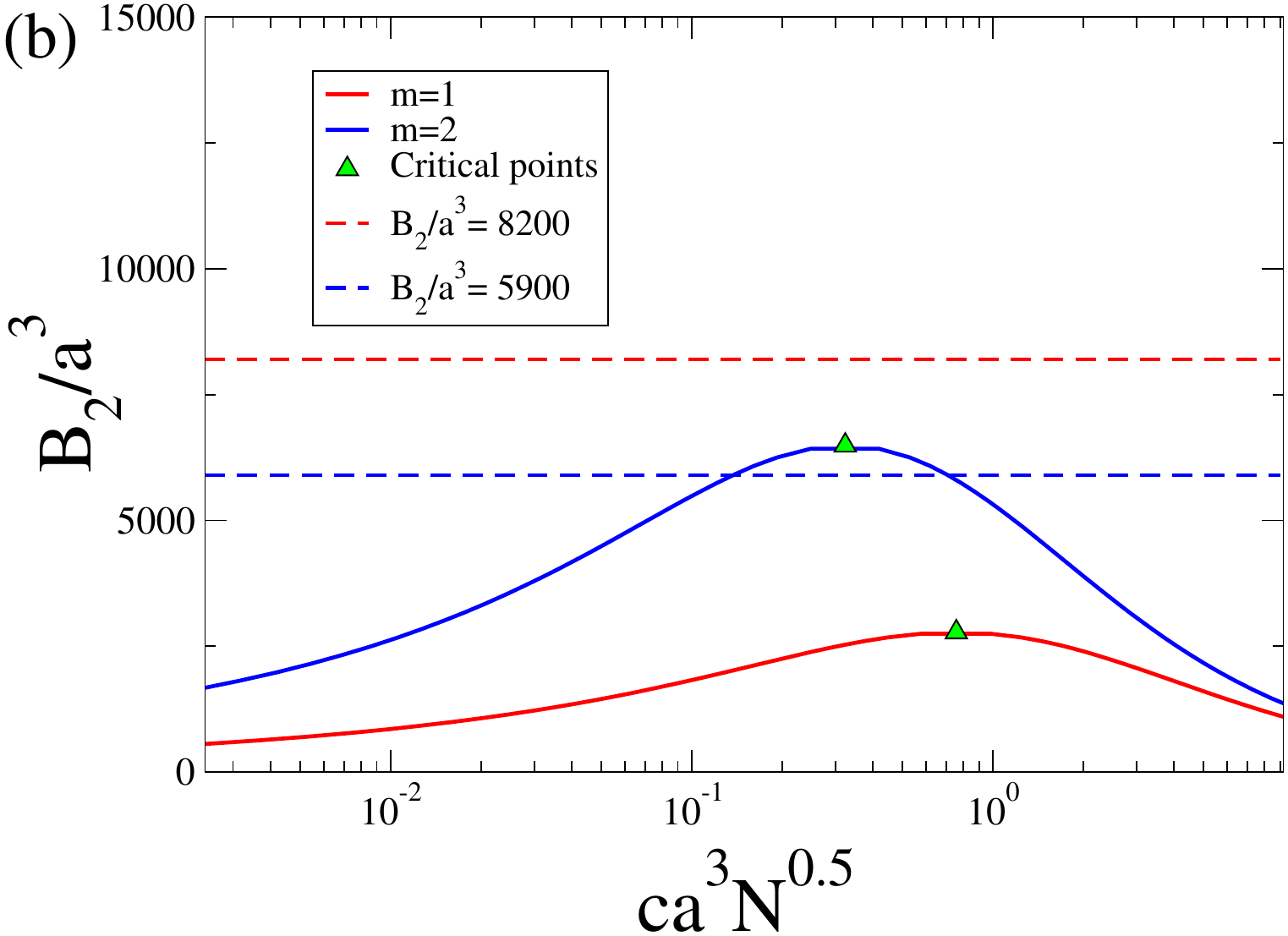}
    \caption{(a) Binodal lines for the $m=1$ and the $m=2$ system of a solution of ideal SCNPs. Here $\mathcal{M}=24$, $a=1$, $l=10$ and $N=\mathcal{M}l=240$. (b) Binodal lines for the $m=1$ and $m=2$ system of a solution of real SCNPs, for the same parameter values. The dashed lines, which match the corresponding system colour, indicate the value of the chain-chain second virial coefficient, as extracted from simulations (see the {\it Methods}). In both panels the coexistence and critical points have been computed numerically by solving Eqs.~\eqref{eq:coexistence} and~\eqref{eq:criticality}, respectively.} 
    \label{fig:phase_diag}      
\end{figure}

After numerically solving the relevant systems of equations, Eqs~\eqref{eq:coexistence} and~\eqref{eq:criticality}, we find that, for a given set of $\mathcal{M}$, $l$ and $m$, there is a critical value of the chain-chain second virial coefficient, $B_{2,m}^{\rm crit}$, above which the system never phase-separates. Figure~\ref{fig:phase_diag} shows examples for two systems (respectively with $m=1$ and $m=2$) previously investigated numerically, for both ideal, Fig.~\ref{fig:phase_diag}(a), and real, Fig.~\ref{fig:phase_diag}(b), chains. In the latter case, we also show the values of the $B_2$ evaluated in simulations for the two types of chains as horizontal dashed lines.
 The figure shows that, consistently with the numerical findings~\cite{rovigatti2022designing}, the theory predicts that the system with $m=1$ does not phase-separate, while $m = 2$ real chains do.

 Notably, for the case in which phase separation is present ($m=2$) the theory predicts a value for the concentration of the dense phase  $c^{\rm dense}a^3 \sqrt{N} \approx 1$ which is in reasonable agreement with the numerical value of $\approx 1.6$~\cite{rovigatti2023}.

\section{Scaling of the critical parameters}

In order to analytically characterise the scaling properties of the critical parameters of the system, we simplify Eq.~\eqref{q} by assuming that $c\frac{v_L}{L} \ll 1$ for $c$ close to $c_{\rm crit}$, which, as we will show \textit{a posteriori}, is true if $\mathcal{M}$ is sufficiently large. Since $v_L\sim a^3 L^{3\nu}$,  the former assumption implies $c a^3 L^{3\nu-1} \ll 1$. We can now write:
\begin{equation}
q\approx  \frac{4cv_L}{2^{3\nu+1}L}
\end{equation}

\noindent
with $q\ll 1$. As a result, only few stickers are involved in intermolecular bonds, which confirms that the bonding part of the free energy can be treated as a perturbation of the reference state, where only intra-molecular bonds are present. This is coherent with numerical results, which show that close to the critical point the fraction of stickers involved in intermolecular bonds in the high-density phase is $\approx 0.12$~\cite{rovigatti2023}. Because such a small fraction of stickers forms intermolecular bonds, the vast majority of stickers on each chain are engaged in intra-chain bonding and thus unavailable for inter-chain interactions. This suppresses correlations between stickers belonging to the same chain: a free sticker is essentially equally (un)likely to encounter another free sticker from its own chain as from a different one. Consequently, all available stickers become effectively indistinguishable, thereby justifying the use of a mean-field description.

In the aforementioned low $q$ approximation, the criticality conditions become:

\begin{equation}
\begin{cases}
\frac{2B_2}{N^2}+\frac{1}{cN}-m\frac{4a^3L^{3\nu-2}}{2^{3\nu+2}}+3m\frac{16 a^6L^{6\nu-3}}{2^{6\nu+3}}c &= 0\\
-\frac{1}{c^2 N} + 3m\frac{16 a^6L^{6\nu-3}}{2^{6\nu+3}} & = 0
\end{cases}
\end{equation}

\noindent
which can be solved to yield

\begin{equation}\label{crit_c_2}
c^{\rm crit}_m=\frac{2^{3\nu+\frac{3}{2}}\mathcal{M}^{3\nu-\frac{3}{2}}}{4\sqrt{3}m^{3\nu-1} N^{3\nu-1}a^3}
\end{equation}

\begin{equation}\label{eq:B2_crit}
B^{\rm crit}_{2,m}=\left(\frac{m}{2}\right)^{3\nu -1}\frac{a^3N^{3\nu}}{\mathcal{M}^{3\nu-\frac{3}{2}}}\left(\frac{\mathcal{M}^{\frac{1}{2}}}{2^{2}}-\frac{\sqrt{3}}{2^{\frac{1}{2}}}\right)
\end{equation}

\begin{figure}
    \centering
    \includegraphics[width=0.6\linewidth]{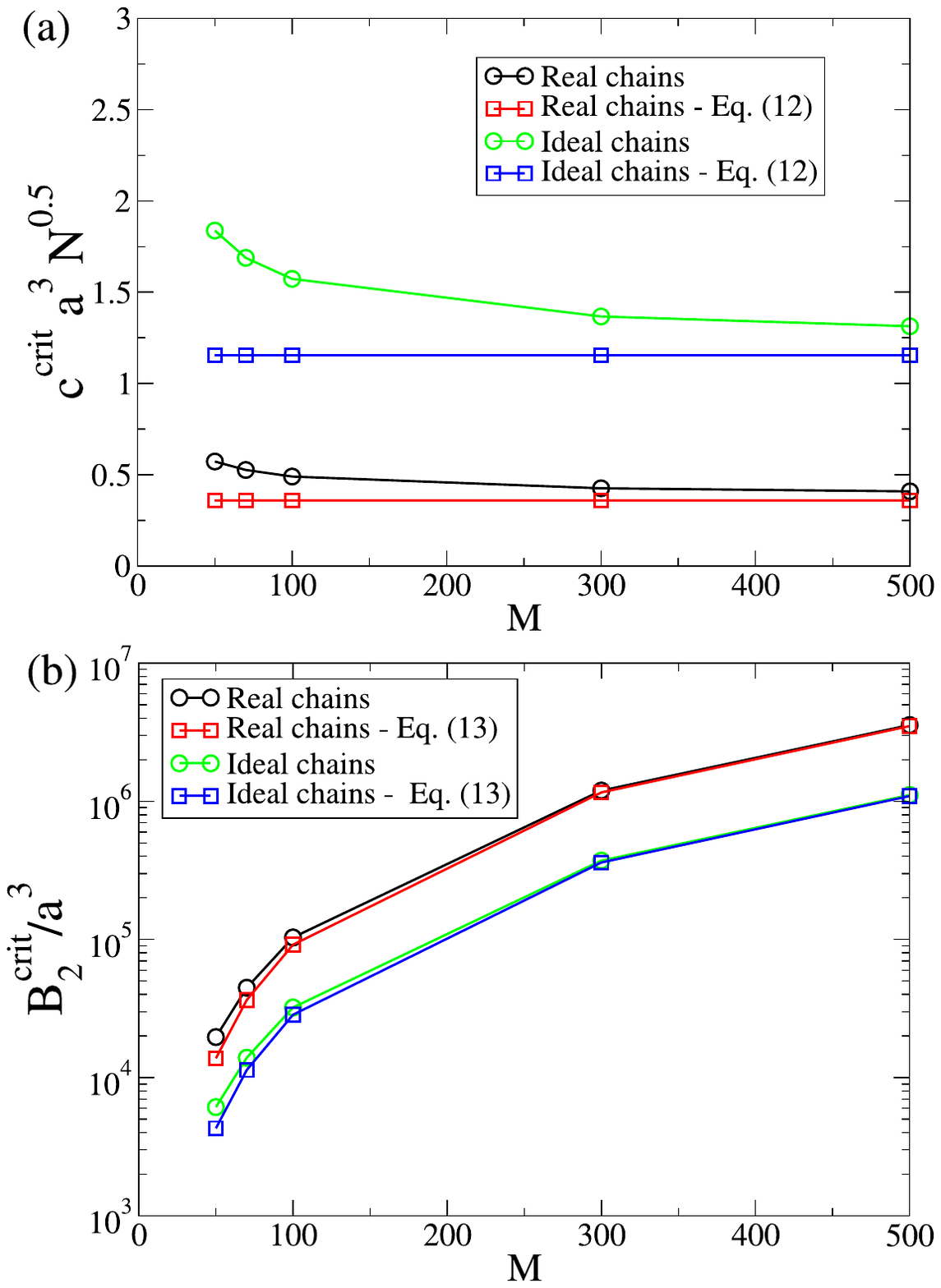}
    \caption{(a) Comparison between the numerical estimates of the critical monomer density (obtained solving system~\eqref{eq:criticality}) and their low $q$ approximation from Eq.~\eqref{crit_c_2}, as a function of the number of stickers $\mathcal{M}$ at $l=10$ and $m=1$, for both ideal and real chains. (b) Comparison between the numerical estimates of the critical excluded volume (obtained solving system~\eqref{eq:criticality}) and their low $q$ approximation from Eq.~\eqref{eq:B2_crit}, as a function of $\mathcal{M}$ at $l=10$ and $m=1$, for both ideal and real chains. In both cases the low-$q$ approximation improves as $\mathcal{M}$ increases.} 
    \label{fig:panels_approx}      
\end{figure}

By using Eq.~\eqref{crit_c_2} we can check that the assumption $\frac{c^{\rm crit}_mv_L}{L}\sim \mathcal{M}^{-\frac{1}{2}} \ll 1 $ holds near the critical point at fixed values of $\mathcal{M}$ and $L$ when $\mathcal{M} \gg 1$, confirming that, in this limit, $q \ll 1$. Panels (a) and (b) of Fig.~\ref{fig:panels_approx} show the agreement between the numerical solutions of system~\eqref{eq:criticality} and the approximated expressions of the critical point from Eqs.~\eqref{crit_c_2} -~\eqref{eq:B2_crit} at low $q$, as a function of the number of stickers $\mathcal{M}$. For both $c^{\rm crit}_m$ and $B^{\rm crit}_{2,m}$ the agreement increases with $\mathcal{M}$.

Eq.~\eqref{eq:B2_crit} can be used to rationalise the available numerical data~\cite{rovigatti2022designing,rovigatti2023}, since it shows that increasing $m$ explicitly increases the tendency to phase-separate. Indeed, at fixed $l$, $\mathcal{M}$, and $N = l \mathcal{M}$, the critical second virial coefficient scales as $B^{\rm{crit}}_{2,m} \sim m^{3\nu-1}$.

It is also interesting to look at how $B_{2, m}^{\rm crit}$ scales when $\mathcal{M}$ and $m$ are kept fixed, \textit{i.e.} when $l$ (and therefore $N$) changes. In this case, for Gaussian chains $B^{\rm crit}_{2,m}\sim N^{\frac{3}{2}}$, while for real chains $B^{\rm crit}_{2,m}\sim N^{2.225}$. If we assume that in both cases the second virial coefficients scale with the gyration radius $R_g$ as $B_{2, m} \sim R_g^3 \sim N^{\frac{3}{2}}$ (as demonstrated for real SCNPs in a good solvent in both experiments~\cite{ruiz2025} and simulations~\cite{moreno2018}), we find that $\frac{B^{\rm crit}_{2,m}}{B_{2,m}}$ is independent of $l$ for Gaussian chains, while scales as $\sim l^{0.725}$ for real chains. Therefore, for Gaussian chains the theory predicts that the phase behaviour of a system of associative polymers with $\mathcal{M}$ stickers of $m$ types is not affected by $l$. By contrast, for real chains the driving force for phase separation increases with $l$, so that a given system will eventually phase-separate, provided that $l$ is large enough. For instance, for the $m=1$ case with $\mathcal{M} = 24$, which does not separate in numerical simulations when $l = 10$, our theory predicts that $\frac{B^{\rm crit}_{2,m}}{B_2} \geq 1$, and therefore the system can undergo phase separation, for $l \gtrsim 45$, a value for which simulations are currently unfeasible. For $m=2$, we find that $\frac{B^{\rm{crit}}_{2, m}}{B_{2,m}}\geq 1$ for $l\gtrsim 9$, a value coherent with available numerical simulations and with our theory, while systems with $m=4$ (which have a strong tendency to phase separate and for which numerical data to compute $B_{2, m}$ is also available, see Fig.~\ref{fig:betaV}) display phase separation for $l \gtrsim 3$. In the future, it would be interesting to test which version of the theory better describes the behaviour of SCNPs in a good solvent, both via simulations and experiments.

\section{Conclusions}
In conclusion, we have constructed a mean-field theory able to capture the physics of phase separation in associative polymers decorated with alternating stickers of $m$ different types, corroborating numerical results showing how increasing $m$ enhances the driving force for phase separation. We have shown that a virial expansion at the chain level (\textit{i.e.} which uses the full-chain excluded volume instead of the monomer excluded volume) for the repulsive term in the free energy is effective at reproducing the equation of state obtained from numerical simulations of SCNPs. Indeed, notwithstanding the approximations, the theory matches numerical results at least semi-quantitatively, correctly predicting the presence of phase separation when $m=2$. 

Overall, our results provide a compact, yet physically transparent framework for understanding how the number and type of stickers control the phase behaviour of SCNP solutions. Forming bonds between different polymers  unfreezes some of the entropy 
captured in the isolated fully-bonded SCNP. The amount of this entropic gain may (or may not, depending on the number of sticker types) be sufficient to drive a macroscopic phase separation process. Despite its simplicity, the theory identifies the dominant entropic mechanisms and highlights which microscopic assumptions (looping statistics, chain-level excluded volume) are most relevant for accurately capturing the thermodynamics of these systems.

We note that our approach can be extended to more general systems, such as polymer chains decorated with non-equispaced stickers along the backbone. Since such systems play important roles in \textit{e.g.} biological settings~\cite{rasid_2021, feldman_2009}, we plan to tackle the investigation of such disordered systems with theory and simulation in the future.

\section{Methods}\label{sect:methods}

\subsection{Phase diagram calculation}
To compute the binodal lines in Fig.~\ref{fig:phase_diag}, we have numerically solved the following system of equations enforcing chemical and mechanical equilibrium between the polymer-poor (gas) and polymer-rich (liquid) phases for a given value of the second virial coefficient $B_2$:
\begin{equation}\label{eq:coexistence}
\begin{cases}
\mu(c^{\rm{liquid}}, B_2)=\mu(c^{\rm{gas}}, B_2)\\
P(c^{\rm{liquid}}, B_2)=P(c^{\rm{gas}}, B_2)\\
\end{cases}
\end{equation}

\noindent
where $\mu(c, B_2)=\frac{\partial f(c, B_2)}{\partial c}$ is the chemical potential and $P(c, B_2)=c\frac{\partial f(c, B_2)}{\partial c}-f(c, B_2)$ is the associated pressure.

The critical point $(c^{\rm{crit}}, B_2^{\rm{crit}})$ has been estimated by numerically finding the maximum of the spinodal line $\frac{\partial^2f(c, B_2)}{\partial c^2}=0$, for which the following conditions hold:
\begin{equation}\label{eq:criticality}
\begin{cases}
\left.\frac{\partial^2 f(c, B_2)}{\partial c^2}\right|_{c^{\rm{crit}}, B_2^{\rm{crit}}}=0\\
\left.\frac{\partial^3 f(c, B_2)}{\partial c^3}\right|_{c^{\rm{crit}}, B_2^{\rm{crit}}}=0
\end{cases}
\end{equation}

\subsection{Molecular simulations}

We have simulated polymer chains in the $NVT$ ensemble by coupling the system to an Andersen-like stochastic thermostat~\cite{russo2009}. The interaction between two monomers sharing a covalent bond, \textit{i.e.} between two bonded neighbours, is given by the Kremer-Grest potential~\cite{kremer1990}:
\begin{equation}
U_{\rm bond}(r)=U_{\rm FENE}(r)+U_{\rm WCA}(r)
\end{equation}
with $r$ the distance between the two interacting monomers. The attractive, finitely extensible nonlinear elastic term is:
\begin{equation}
U_{\rm FENE}(r)=
\begin{cases}
-\dfrac{1}{2} K_0 r_0^2 \log{\left(1-\left(\dfrac{r}{r_0}\right)^2\right)} & r<r_0\\[6pt]
+\infty & r>r_0
\end{cases}
\end{equation}
with $K_0 = 30\frac{ \epsilon_{\rm LJ}}{\sigma^2}$ the bond stiffness and $r_0 = 1.5\sigma$ the maximum extension (where $\sigma$ and $\epsilon_{\rm LJ}$ define the length and energy units).  
The excluded volume contribution is given by the purely repulsive WCA potential~\cite{weeks1971}:
\begin{equation}
\label{eq:WCA}
U_{\rm WCA}(r)=
\begin{cases}
4\epsilon_{\rm LJ}\!\left[\left(\dfrac{\sigma}{r}\right)^{12} - \left(\dfrac{\sigma}{r}\right)^6\right] & r<2^{1/6}\sigma\\[6pt]
0 & r>2^{1/6}\sigma
\end{cases}
\end{equation}
We adopt reduced units and set $\sigma = m = \epsilon_{\rm LJ} = 1$. The potential is truncated at $2^{1/6}\sigma$ to remove the attractive contribution.

The interaction between non-bonded neighbours is given by a WCA repulsion (present only for non-ideal chains), Eq.~\eqref{eq:WCA}, and by an additional short-range attractive term that acts only between stickers of the same type. The latter potential is modelled after the Stillinger-Weber potential~\cite{stillinger1985}, widely adopted in soft-matter models of short-range attraction~\cite{gado2007, saw2009}:
\begin{equation}
U_{\rm sticker}(r)=
\begin{cases}
C\epsilon\,[D(\sigma_s/r)^4 - 1] \exp\!\left(\dfrac{\sigma_s}{r-r_c}\right) & r<r_c\\[6pt]
0 & r>r_c
\end{cases}
\end{equation}
with $\sigma_s = 1.05\sigma$, $r_c = 1.68\sigma$, $C=8.97$, $D=0.41$. The attraction depth  (the energy scale of the bond interaction) is $\epsilon$.

To enforce the one-bond-per-sticker condition, we include a three-body potential $U_{\rm 3b}$ that suppresses the formation of bonded triplets~\cite{sciortino2017}. Specifically, $U_{\rm 3b}$ introduces a repulsive contribution that compensates the energy gained upon forming a second bond. Thus, when monomer $i$ approaches a bonded pair $(j,k)$, it moves along an almost flat energy hypersurface, meaning that configurations where $i$ binds to either $j$ or $k$ (while the third monomer is free to detach) are not separated by any energy barrier.

The three-body potential reads:
\begin{equation}
U_{\rm 3b}(r_{ij}, r_{ik})
= \epsilon \sum_{i,j,k} U_{\rm 3}(r_{ij})\, U_{\rm 3}(r_{ik})
\end{equation}
where the sum runs over all bonded triplets, defined as configurations in which particle $i$ is within $r_c$ from both $j$ and $k$.

The two-body factor $U_{\rm 3}(r)$ is:
\begin{equation}
U_{\rm 3}(r) =
\begin{cases}
1 & r \le \sigma_s\\[6pt]
-\dfrac{U_{\rm sticker}(r)}{\epsilon} & r>\sigma_s
\end{cases}
\end{equation}
where $\sigma_s$ is the position of the minimum of $U_{\rm sticker}$.

\begin{figure}
    \centering
    \includegraphics[width=0.6\linewidth]{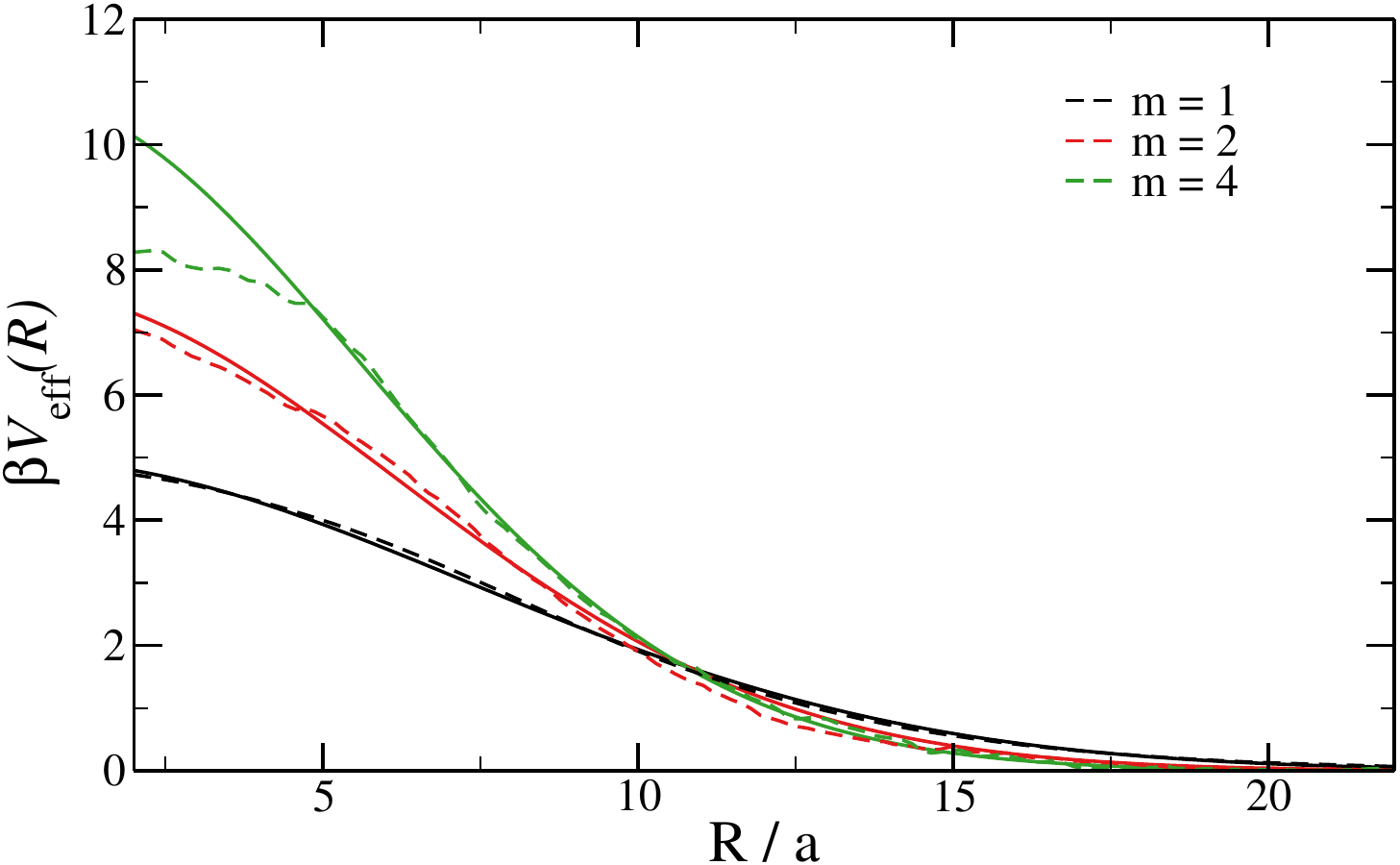}
    \caption{Dashed lines are the effective interactions between fully-bonded SCNPs with $\mathcal{M} = 24$, $l = 10$, and three types of sticker sequence, $m = 1$, $m = 2$ and $m=4$, as a function of the chain-chain centre of mass separation distance $R$, when only intra-chain bonding is allowed. Solid lines are best fits to Gaussian functions, which are then used in Eq.~\eqref{eq:B2} to compute the second virial coefficients, yielding $B_2 = 8200\, a^3$, $B_2 = 5900\, a^3$ and $B_2 = 5500\, a^3$ for the $m = 1$, $m = 2$ and $m=4$ systems, respectively. We note that the discrepancy between the numerical data and the fit for the $m = 4$ case has a tiny effect on the value of the $B_2$ (a few percent, which is smaller than the statistical uncertainty due to the numerical sampling).}
    \label{fig:betaV}
\end{figure}

We have performed the following types of simulations:

\begin{itemize}
    \item \textbf{Simulations of two $A_{i1}$ units}: two sequences of 20 monomers each (both ideal-chain and real-chain conditions) to reconstruct the mass--action law of Fig.~\ref{fig:mass_action}. Each chain contains 2 stickers of the same type separated by 18 inert monomers. Identifying $q$ with the fraction of fully-bonded configurations in which the two chains are connected to each other via inter-chain bonds, we have computed $\langle q/(1-q)^2 \rangle$ at $\beta \epsilon=6$ and different values of the monomer concentration $c$ by varying the simulation box size.
    
    \item \textbf{$B_2$ calculation}: two fully-bonded chains of valence $\mathcal{M}=24$ and sticker-sticker spacing $l=10$, used to compute the excluded volume $B_2$ for a pair of SCNPs. These simulations have been run at $\beta \epsilon = 14$. To calculate the second virial coefficient, we start computing the effective interaction between the two fully-bonded chains, with the constraint that only intramolecular bonding is possible. Using the same methodology described in Ref.~\cite{rovigatti2022designing}, we have evaluated the effective potential $V_{\rm eff}(R)$, where $R$ is the distance between the two centres of mass, for the $m = 1$, $m = 2$ and $m=4$ cases. Fig.~\ref{fig:betaV} shows that the resulting effective potential can be fitted to a Gaussian, from which we compute the second virial coefficient, defined as:

\begin{equation}
\label{eq:B2}
B_2 = -2\pi \int_0^\infty \left[\exp(-\beta V_{\rm eff}(R)) - 1\right] R^2 dR
\end{equation}

whence we obtain $B_2 \approx 8200 \, a^3$ for $m = 1$, $B_2 \approx 5900 \, a^3$ for $m = 2$ and $B_2 \approx 5500 \, a^3$ for $m=4$. 
    
    \item \textbf{Bulk simulations}: 100 SCNPs with $\mathcal{M}=24$ and $l = 10$ and $m = 1$ or $m = 2$, used to compute the equation of state at different values of the monomer concentration $c$ in Fig.~\ref{fig:B2_pressure}.
\end{itemize}

Throughout all simulations, the integration time step was set to $\Delta t = 0.003$, expressed in natural units $\sigma\sqrt{\frac{m}{\epsilon_{\rm LJ}}}$.
\appendix

\section{Mass action law derivation}
\label{app:mass_action}

We remind here that $Q_{A_{i1}}$ and $N_{A_{i1}}$ in Eq.~\eqref{eq:mass_action} indicate the partition function and the number of segments of inert monomers forced to loop by the presence of a bond between two $i$-type stickers (see Fig.~\ref{fig:schema}). Similarly, $Q_{A_{i2}}$ and $N_{A_{i2}}$ indicate the canonical partition function and the number of two inter-polymer bonded $A_{i1}$ units. 
The quantity $N_{A_{i1}}$ ranges between $0$ and $\frac{N_i}{2}$, where the latter is the maximum number of possible $A_{i}$ units, corresponding to half of the total number of stickers of type $i$.
Similarly, $0<N_{A_{i2}}< \frac{N_i}{4}$. Then:
\begin{equation}
N_{A_{i1}}=\frac{N_i}{2}(1-q) ~~~~~~~~~~~~~~N_{A_{i2}} =\frac{N_i}{4}q
\end{equation}

For the $A_{i1}$ unit we can write:
\begin{equation}
Q_{A_{i1}}=\frac{V}{V_0}
\end{equation}
where we have explicitly highlighted the centre of mass contribution given by the volume of the system, $V$.
The term $V_0$ is a volume term encoding the contribution to the $A_{i1}$ partition function from all other  internal degrees of freedom.
\begin{figure}
    \centering
    \includegraphics[width=0.4\linewidth]{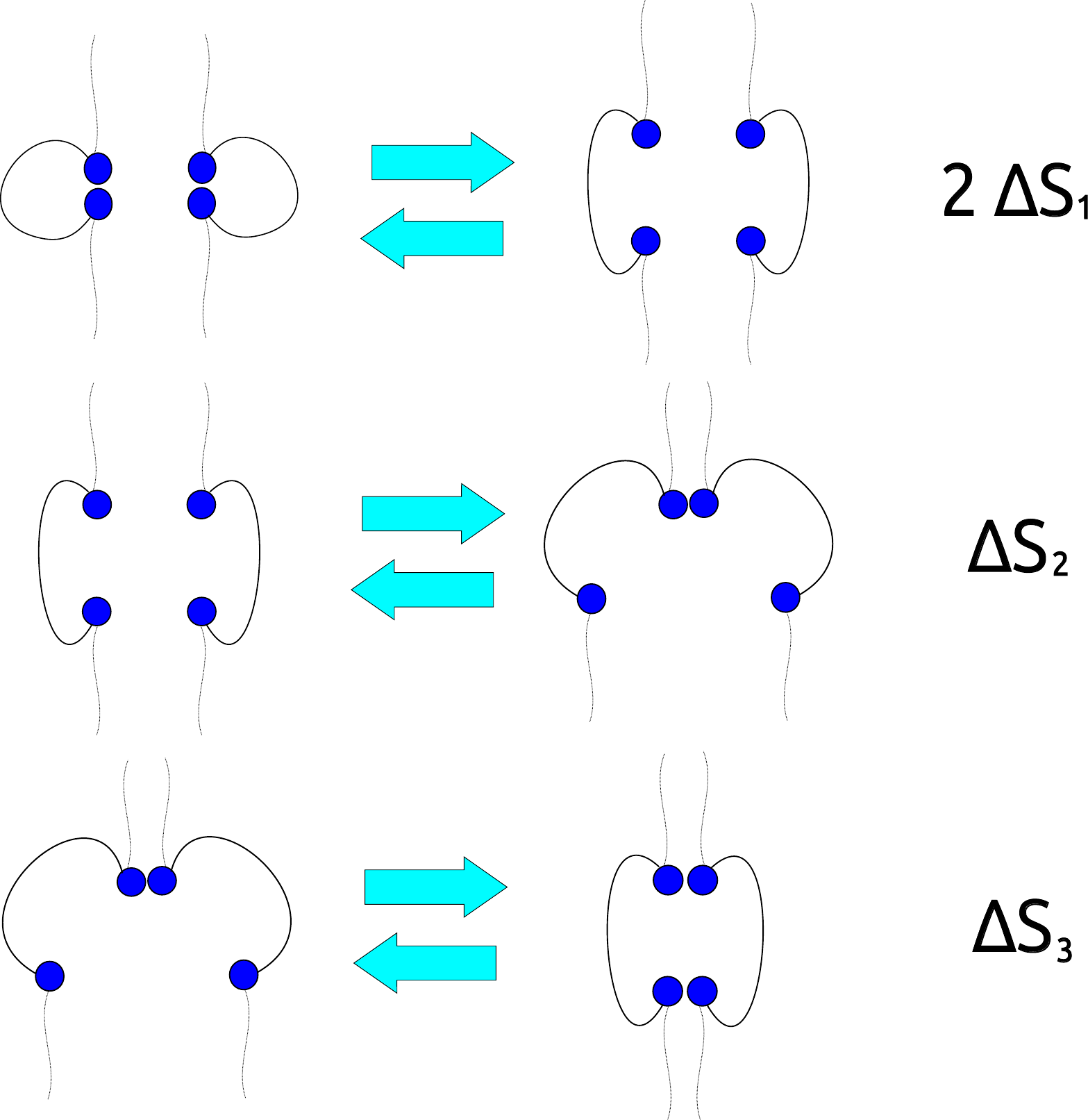}
    \caption{Reaction steps connecting the formation of an $A_{i2}$ complex starting from two $A_{i1}$ units, and associated entropy variations. The entire process is separated in three independent steps, whose entropy contribution is calculated in the text.}
    \label{fig:entropy}
\end{figure}

To work out $Q_{A_{i2}}$, we  start from $Q_{A_{i1}}$ and compute the total entropy variation $\Delta S_{\textrm{tot}}$ arising from breaking the intramolecular bonds of the two interacting units and forming the new intermolecular bonds.

This entropy difference can be decomposed in three different terms, following the steps depicted in Fig.~\ref{fig:entropy}:
\begin{itemize}
    \item the entropy increase $\Delta S_1$ associated with breaking the intramolecular bond between the stickers in a $A_{i1}$ unit to open it. It is such that $e^{\frac{\Delta S_1}{k_B}}=\frac{v_L}{v_b}$; 
    \item the entropy decrease $\Delta S_2$  associated with the formation of a bond between two open chains of length $L$, which generates an open chain of length $2L$. It is such that $e^{\frac{\Delta S_2}{k_B}} =\frac{4 v_b}{V}$, where the factor 4 accounts for the different ways in which two free chains can bond to each other;
    \item the entropy decrease $\Delta S_3$ due to the formation of a closed complex of size $2L$. It is such that $e^{\frac{\Delta S_3}{k_B}}=\frac{v_b}{v_{2L}}=\frac{v_b}{2^{3\nu} v_L}$. Here the factor $2^{3\nu}$ has been added assuming a generic power scaling law $v_L\sim a^3 L^{3\nu}$ for SCNPs (for some references on the effective scaling behaviour of SCNPs, see~\cite{ruiz2025, moreno2018}).
\end{itemize}
\noindent

Overall, the total entropy contribution due to moving from two intramolecular $A_{i1}$ units to a dimer is:

\begin{equation}
\Delta S_{\textrm{tot}}=2\Delta S_1+\Delta S_2 + \Delta S_3=k_B\log\frac{4 v_L^2}{v_{2L}V}=k_B \log{\frac{4v_L}{2^{3\nu}V}}
\end{equation}

This means that the final expression for the dimer partition function amounts to:

\begin{equation}
Q_{A_{i2}}=\frac{(Q_{A_{i1}})^2}{2!}e^{\frac{\Delta S_{\textrm{tot}}}{k_B}}= \frac{(Q_{A_{i1}})^2}{2}\frac{4v_L}{2^{3\nu}V}
\end{equation}

\noindent
where the factor $2!$ counts the number of indistinguishable states of the system: the two $A_{i1}$ in the first step of the transformation in Fig.~\ref{fig:entropy} are identical.

Plugging  these expressions in Eq.~\ref{eq:mass_action}, one gets:

\begin{equation}\label{q1}
\frac{q_i}{(1-q_i)^2}=\frac{4N_{A_{i}} v_L}{2^{3\nu}V}
\end{equation}

\noindent
where $N_{A_{i}}=\frac{N_i}{2}=\frac{N^{\textrm{tot}}}{2L}$, with $N^{\textrm{tot}}$ total number of monomers. Plugging this expression in Eq.~\eqref{q1}, we obtain:

\begin{equation}
\frac{q_i}{(1-q_i)^2}=\frac{4c v_L}{2^{3\nu+1}L}
\end{equation}
with $c=\frac{N^{\textrm{tot}}}{V}$.
 
\section{The free energy of an ideal gas of clusters}
\label{app:free_en_clusters}
We review here the expression for the bonding contribution to the Helmholtz free energy~\cite{hill2012introduction,wertheim1984fluids}, following the derivation proposed in Ref.~\cite{chapman1988phase}. We focus on the free energy $F$ of an ideal gas of $N$ particles in a volume $V$ which can dimerise (particles with only one binding site). We will generalise this to a mixture of different particles later on.
For the one-component case, in the ideal gas limit,
if we call $q$ the probability that one bond is formed, the number of 
particles in monomeric state is $N(1-q)$ and the number of
dimers is $\frac{Nq}{2}$.  In the ideal gas, pressure counts the number of clusters, giving $\beta P V =  N(1-q) + \frac{Nq}{2}  = N- \frac{Nq}{2} $.
Similarly, the chemical potential coincides with the logarithm of the
monomer concentration $\beta \mu=\ln (N(1-q)V_0/V)$, where the term $V_0$ is a reference volume encoding the contribution from all other internal degrees of freedom.

Since the Gibbs free energy is $N\mu$, we have:
\begin{equation}\label{eq:F}
\beta F +\beta PV   = N \beta\mu 
\end{equation} 
and thus:
\begin{equation}
\beta F =  
 N\ln \frac{N V_0}{V} -N  + N \ln(1-q) +\frac{ Nq}{2}
\end{equation}
which can be rewritten as the ideal gas contribution $N(\ln\frac{N V_0}{V}-1)  $ plus a bonding contribution $F_{\rm bond}$
with
\begin{equation}
\beta F_{\rm bond}= N \ln(1-q) + \frac{ Nq}{2}
\end{equation}

For an ideal gas of different particles, the reference free energy includes the mixing term, but the bonding contribution remains identical and needs only to be summed over the different species.
As commonly done~\cite{wertheim1984fluids}, the ideal gas free energy is substituted by more accurate reference free energies.  We do the same here, using as reference free energy the virial expression for 
SCNP.  For our case, the bonding contribution can then be expresses as in Eq.~\ref{sticker_free_en}.

\section{The generalisation of the Semenov-Rubinstein theory to $m > 1$}
\label{app:sem_rub_gen}
We are interested in extending the theory by Semenov and Rubinstein in Ref. [13] to the case of polymer chains with multiple sticker types $m>1$.

We start from the original renormalised version of the Semenov-Rubinstein theory, in which the reference state is a solution of SCNPs with steric interactions encoded in $\beta f_{\textrm{ex}}(c)$ happening at the monomer-monomer level~\cite{semenov1998}. Using a third-order virial expansion in $c$, we get for the reference part of the free energy density:
\begin{equation}
\beta f_{\rm{ref}}(c)=\frac{c}{N}\log{\left(\frac{c}{eN}\right)} + \beta f_{\rm{ex}}(c)
\end{equation}
with $\beta f_{\rm{ex}}(c) = B_2 c^2+B_3 c^3$. Here $B_2$ indicates the excluded volume between two monomers and $B_3$ is the third virial coefficient.

To compute the inter-chain bonding contribution $\beta f_{\rm{bond}}(c)$, we leverage the same mean-field, polymer-science based approach Semenov and Rubinstein use in their theory. Following their original derivation, we first write the inter-chain  bonding partition function:
\begin{equation}
Q_{\rm{bond}}=\prod_{i=1}^m Q_{\rm{bond}}^{(i)}
\end{equation}
where $Q_{\rm{bond}}^{(i)}$ indicates the partition function due to the interaction between stickers of type $i$, of the form:
\begin{equation}
Q_{\rm{bond}}^{(i)}=P^{(i)}_{\rm{comb}}W_i \exp{\left(N^{(i)}_p \frac{\Delta S_i}{k_B}\right)}
\end{equation}
for each $i \in \{1,..., m\}$. Since each sticker can only bond with one other sticker (provided they are of the same type), we have the same partition function for each sticker type as the $m=1$ case. To be more precise, $P^{(i)}_{\rm{comb}}$ is the number of ways of pairing $2N^{(i)}_{\rm{p}}$ stickers out of their total number $N^{(i)}_{\rm{st}}$, $W_i$ is the probability that, in the absence of attractive interactions, all $N^{(i)}_p$ stickers can be found close enough to the partner with which they share a bond, and $\exp{(\frac{\Delta S_i}{k_B})}=\frac{v_L}{v_b}$ is the Boltzmann factor linked to the entropy variation due to the breaking of an intra-chain bond. Then:
\begin{equation}
P_{\rm comb}^{(i)}=
\frac{N_{\rm st}^{(i)}!}{\left(N_{\rm st}^{(i)}-2N_p^{(i)}\right)!\,\left(N_p^{(i)}!\right)\,2^{N_p^{(i)}}}
\end{equation}
And:
\begin{equation}
W_i=\left(\frac{v_b}{V}\right)^{N_p^{(i)}}
\end{equation}
for each type of sticker. Keeping in mind that $\beta f_{\rm{bond}}=-\frac{\log{(Q_{\rm{bond}})}}{V}$ and substituting, we get:
\begin{equation}\label{eq:sr_gem_var}
\begin{aligned}
\beta f_{\rm bond}= \sum_{i=1}^{m}\left(
\frac{c}{2L}\left(q_i\ln q_i+2(1-q_i)\ln(1-q_i)\right) -\frac{c q_i}{2L}\ln\!\left(\frac{c v_L}{e L}\right)
\right)
\end{aligned}
\end{equation}
The next step is to minimise the total free energy density of the system with respect to each degree of conversion $q_i$  to find the equilibrium state. Such a minimisation yields $m$ equationsthat provide the conditions for a minimum free energy:
\begin{equation}
\frac{q_i}{(1-q_i)^2}=\frac{c v_L}{L}
\end{equation}
We note that the attractive volume of a bond is the same as that of the $m=1$ system, since it doesn’t depend on the type of the stickers. Solving these equations gives the value of the degree of conversion $q$:
\begin{equation}
q=1-\frac{\sqrt{1+4c \frac{v_L}{L}}-1}{2c \frac{v_L}{L}}
\end{equation}
Substituting in Eq.~\ref{eq:sr_gem_var}, we finally get:
\begin{equation}
\beta f_{\rm{bond}}=\frac{mc}{L}\left(\frac{ q}{2}+\ln(1-q)\right)
\end{equation}
The full expression of the free energy density in this framework is then:
\begin{equation}\label{eq:free_en_sem_gen}
\beta f(c)=\frac{c}{N}\log{\left(\frac{c}{eN}\right)} + \beta f_{\rm{ex}}(c) + \frac{m c}{L}\left(\frac{q}{2} + \log{(1-q)}\right)
\end{equation}
which bears a significant similarity with the free energy density of Eq.~\eqref{free_energy} in the main text. We note that, using the excess free energy of Ref.~\cite{semenov1998} in a good solvent, this extension of the theory predicts the absence of phase separation for physically reasonable values of $m$.

\section*{Acknowledgements}

We thank Sanat K. Kumar for helpful discussions. We acknowledge support from Cineca-ISCRA for HPC resources.
L.R. acknowledges support from MUR-PRIN Grants No. 20225NPY8P and P2022JZEJR, funded by European Union ``NextGenerationEU'', Missione 4 Componente 2-CUP B53D23028570001.
\\
\\
\section*{References}
\bibliography{main}

\end{document}